\documentclass{PoS}

\usepackage{mciteplus}

\title{Large $N_c$ Thermodynamics with Dynamical Fermions}

\ShortTitle{Large $N_c$ Thermodynamics with Dynamical Fermions}

\author{Thomas DeGrand\\
        University of Colorado Boulder, Boulder, CO USA
}
\author{\speaker{Daniel C. Hackett}\\
        University of Colorado Boulder, Boulder, CO USA\\
        E-mail: daniel.hackett@colorado.edu
}
\author{Ethan T. Neil\\
	University of Colorado Boulder, Boulder, CO USA\\
	RIKEN-BNL Research Center, Brookhaven National Laboratory, Upton, NY USA
}
        
\abstract{
We present a progress report on our investigation of the thermodynamics of QCD with $N_f=2$ flavors of dynamical Wilson fermions in the limit of a large number of colors $N_c$.
To date, studies of the thermodynamics of QCD at large $N_c$ have been limited to the quenched approximation, i.e., to the behavior of pure $\mathrm{SU}(N_c)$ gauge theory at large $N_c$.
This is the first study of thermodynamics at large $N_c$ using dynamical fermions, and thus the first study able to test whether the quenched approximation is a valid way to investigate large $N_c$ thermodynamics.
After reviewing 't Hooft's large $N_c$ limit, we discuss the automation we use to make this study feasible, and finally compare our preliminary physics results for $\mathrm{SU}(3-5)$ with large $N_c$ expectations.
}

\FullConference{The 36th Annual International Symposium on Lattice Field Theory - LATTICE2018\\
		22-28 July, 2018\\
		Michigan State University, East Lansing, Michigan, USA.}

\begin{document}

\section{Large $N_c$ and thermodynamics}

The large $N_c$ expansion, as originally proposed by 't Hooft \cite{tHooft:1973alw}, offers a controlled, analytic description of the nonperturbative dynamics of QCD.
Large $N_c$ reasoning asserts that when one allows the number of colors $N_c$ to vary while holding the 't Hooft coupling $\lambda=g^2 N_c$ constant, any observable has a characteristic scaling with $N_c$ up to $1/N_c$ corrections.
The coefficients of this expansion contain nonperturbative physics, but the scaling with $N_c$ can be determined using simple diagrammatic counting rules.
This ansatz has been confirmed in an extensive literature of lattice studies which found that many spectroscopic observables, thermodynamic observables, and simple matrix elements scale with $N_c$ as expected \cite{Lucini:2012gg}.

The value of the large $N_c$ approach is that it can give us intuition about \emph{why} QCD behaves the way it does.
Taking the 't Hooft limit $N_c \rightarrow \infty$, QCD simplifies and analytic predictions become possible.
For example, in the large-$N_c$ limit, mesons are infinitely narrow, and the quark model and the OZI rule become exact statements \cite{Witten:1979kh}.
This resembles nature, where mesons are long-lived and the quark model and OZI rule are useful but approximate.

However, thermodynamics at large $N_c$ does not resemble the thermodynamics of QCD.
Because there are $\sim N_c^2$ gluonic degrees of freedom but only $N_c$ fermionic ones, as $N_c$ becomes large the gluons dominate the dynamics and the fermions ``quench out''.  Thus, the large-$N_c$ behavior of QCD is that of pure gauge theory.
The $N_c \rightarrow \infty$ limit of quenched QCD exhibits a first-order transition at $T_c \sim 260$~MeV \cite{Lucini:2012wq}, while real-world $N_c=3$ QCD exhibits a crossover around $T_c \sim 155$~MeV \cite{Soltz:2015ula}.
Is this discrepancy a failure of large-$N_c$ reasoning, or are we being somehow naive in taking the $N_c\rightarrow \infty$ limit?

We need to include dynamical fermions in order to test whether large $N_c$ QCD behaves thermodynamically like pure gauge theory.
This motivates our present study: to the best of our knowledge, all studies of thermodynamics at large $N_c$ thus far have been performed in the quenched approximation, which \emph{assumes} that pure gauge theory is the correct limit of QCD.
In fact, with only one exception known to us \cite{DeGrand:2016pur}, all studies of large $N_c$ phenomena thus far have used the quenched approximation.
Whether this is in general a valid approach to study large $N_c$ QCD ought to be tested, and thermodynamics is a particularly clean testbed: 
large $N_c$ provides several qualitative predictions for thermodynamics, to be discussed below.

We expect the computational cost to scale as roughly $N_c^3$ holding all else fixed, so we limit this initial investigation to $N_c=3,4,5$ and leave higher $N_c$ for future work.
We exclude $\mathrm{SU}(2)$ from our study because both the chiral symmetry breaking pattern in the chiral limit and the order of the deconfinement transition in the pure gauge limit are different than for $N_c \ge 3$.
At present we have only investigated $N_f=2$.
Our results below include data from only $12^3\times6$ lattices.

\section{Numerical Methods}
\label{sec:methods}

\subsection{Data Generation}

We generate all data using variants of MILC adapted to run arbitrary $N_c$.
We use hybrid Monte Carlo (HMC) to generate gauge configurations.
For the gauge sector, we use the unimproved Wilson action.
For the fermions, we use the clover-improved Wilson action constructed from nHYP-smeared fat links with the usual choice of smearing coefficients \cite{Hasenfratz:2001hp, *Hasenfratz:2007rf}.
We set the clover coefficient $c_{sw}=1$, a choice known to work well with nHYP smearing \cite{Shamir:2010cq}.

The simulations are completely ordinary.
We compute screening masses by fitting spatial-direction correlators.
We compute spectroscopic observables using valence compound boundary conditions (also known as the ``periodic plus antiperiodic'' or ``P+A'' trick) to effectively double the spatial extent of our lattices \cite{Blum:2001xb, *Aoki:2005ga, *DeGrand:2007tx}.
All quark masses in the following discussion are the axial Ward identity (AWI) quark mass, computed in the usual way.
The axial Ward identity is insensitive to finite volume effects, so we use the AWI quark mass computed on finite-temperature lattices to estimate its zero-temperature value.
We compute gradient-flowed observables using the unimproved Wilson flow.
In this work we show only preliminary results.

\subsection{Automation}
\label{sec:auto}

Exploring enough Wilson phase diagrams to study thermodynamics at large $N_c$ is a formidable task.
Every distinct combination of $(N_c, N_s, N_t)$ corresponds to a different phase diagram, and we must vary all of these to get control over the chiral, continuum, and $N_c \rightarrow \infty$ limits.
Even looking at only $12^3\times6$ for three values of $N_c$, we have already generated and analyzed data for $321$ independent ensembles, with a typical ensemble having $\sim 60$ configurations.
Automation is necessary to render this study logistically tractable.
To this end, we have constructed a fully-automated generation and analysis pipeline capable of exploring Wilson phase diagrams with minimal human intervention.
Up to a handful of exploratory ensembles run by hand and tuning system metaparameters, all data used in this study have been generated automatically.

A closed loop of three distinct components comprise our system.
First, a lattice-specific scientific workflow management system (WMS) \cite{taxi} coordinates the execution of HMC, spectroscopy, and gradient flow binaries.
The WMS enables automatic detection of common failures, such as too-low acceptance rates and unitarity violations, and allows for limited automatic recovery from such errors.
For further discussion, see Ref.~\cite{Ayyar:2018wwf}.
Second, raw data in output files is parsed and loaded in to a relational (PostgreSQL) database.
With our data centrally organized and rigidly structured, we can automatically maintain an analysis which starts from raw data and yields useful ensemble-level observables like quark masses and phase diagnostics.
Third, an automated phase diagram explorer (APDE) reads the end products of the analysis from the database.
Armed with our current best understanding of each phase diagram, the APDE considers a grid over potentially-interesting parameters $(\beta, \kappa)$ and applies simple criteria
[e.g.: $m_q > 0$, $m_q$ not too large, not too deep in confined or deconfined phase, nearby enough to an already-equilibrated ensemble to reduce equilibration costs]
to determine which grid points are interesting and explorable.
The APDE then specifies new simulation points and directs the WMS to run those simulations,  closing the automation loop.

We use the Polyakov loop at long flow time to diagnose the thermodynamic phase of our ensembles, a choice which is particularly convenient for automation \cite{Ayyar:2017vsu}.
We take an ensemble to be ``confined'' if $| \langle P \rangle |/N_c < 0.25$, ``deconfined'' if $| \langle P \rangle |/N_c > 0.75$, and ``ambiguous'' otherwise, all at flow time $t/a^2=2$.
This is not intended to be a replacement for more quantitative determinations of the location of the (pseudo)critical curve, such as determining the location of the peak in some susceptibility.
However, this approach is sufficient to be used to generate data in the neighborhood of the transition to be used in later, more precise analyses.

\section{Preliminary Results}
\label{sec:results}

\begin{figure}[htp]
	\includegraphics[width=\textwidth]{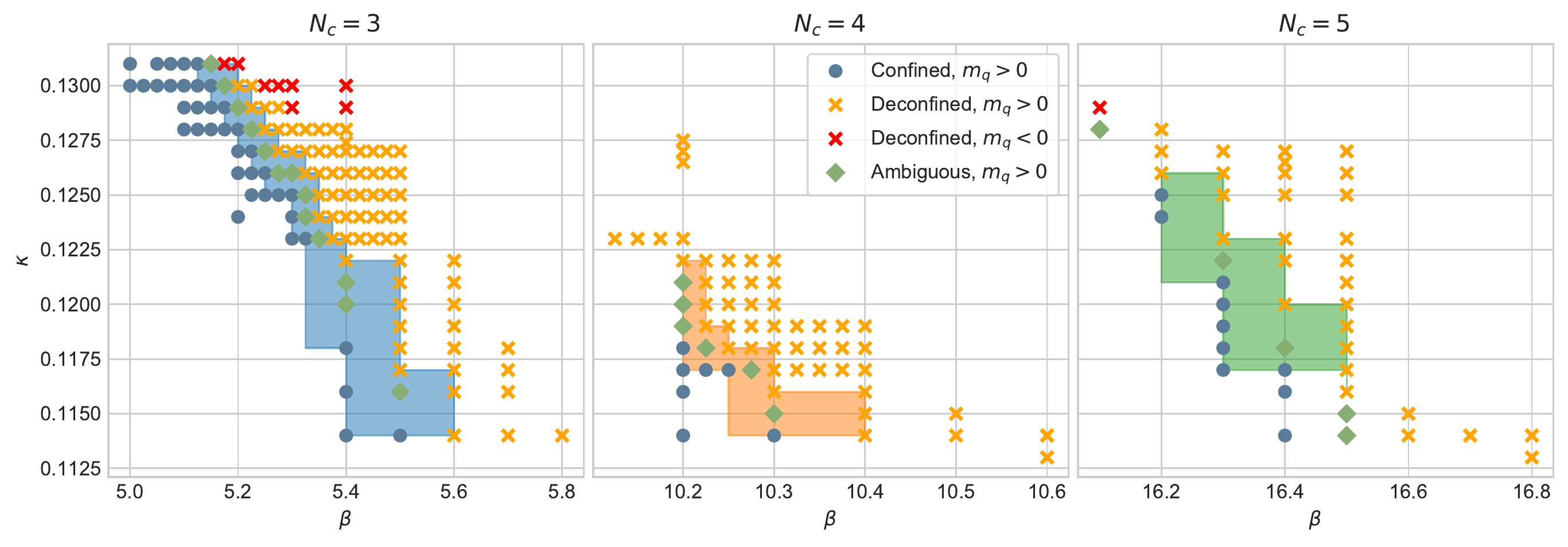}
	\caption{
		Phase diagrams for $N_c=3,4,5$.
		Blue dots are confined ensembles, yellow Xs are deconfined ensembles, and green diamonds are phase-ambiguous ensembles, with all phases diagnosed using the Polyakov loop at long flow time.
		The colored band in each diagram covers the phase-ambiguous region where it is well-determined, with the color corresponding to $N_c$ as in Fig.~\ref{fig:dyn-diag-collapse}.
	}
	\label{fig:phase-diagrams}
\end{figure}

Figure~\ref{fig:phase-diagrams} shows our current results for phase diagrams for $N_c=3,4,5$ with $N_f=2$ flavors on $12^3\times6$ lattices.
For $N_c=4$ and $5$, we are blocked from exploring the transition at lighter quark masses (i.e., ``northwest'') by an apparent bulk transition, to be investigated further in future work.

\subsection{Phase Diagram Collapse}

\begin{figure}[htp]
	\center
	\includegraphics[width=\textwidth]{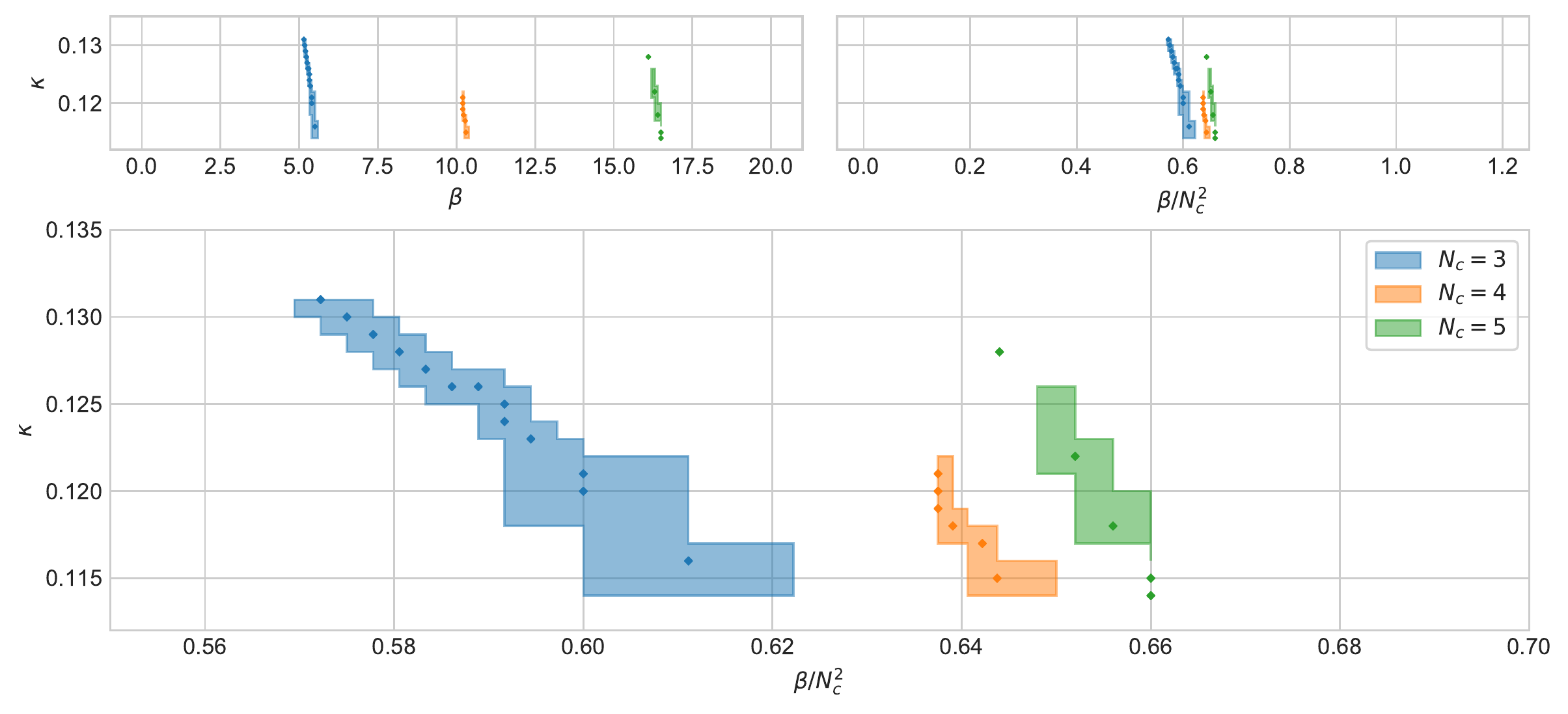}
	\caption{
		Locations of thermal transitions in bare parameter space for $N_c=3,4,5$, plotted together versus $\beta$ (top left panel) and $\beta/N_c^2$ (top right and bottom panels).
		Colored dots are ambiguously-phased ensembles, as determined by Polyakov loops at long flow time.
		Colored bands are phase-ambiguous regions (same as the same-colored bands in Fig.~\ref{fig:phase-diagrams}).
		Together, these provide a rough estimate of the location of the thermal transition.
		When plotted versus $\beta/N_c^2$, the diagrams collapse on top of one another, up to $1/N_c$ corrections.
	}
	\label{fig:dyn-diag-collapse}
\end{figure}

For the large-$N_c$ world to resemble the $N_c=3$ world, physics must remain constant at constant 't Hooft coupling $\lambda \equiv g^2 N_c$, up to $1/N_c$ corrections.
Meanwhile, the fermionic part of the action does not depend explicitly on $N_c$, and so we expect that we do not need to adjust $m_q$ to keep leading-order physics constant as we vary $N_c$.
Translating to lattice couplings, we expect to see constant physics at constant $(\beta/N_c^2, \kappa)$ up to $1/N_c$ corrections.
Thus, we expect Wilson phase diagrams for different $N_c$ to collapse on top of one another when plotted in terms of these variables.

As illustrated in figure~\ref{fig:dyn-diag-collapse}, this is indeed what we observe.
At top left, the transition curves for different $N_c$ are widely separated when plotted versus $(\beta, \kappa)$.
At top right, plotting instead versus $(\beta/N_c^2, \kappa)$, we see the transition lines cluster together.
The bottom panel zooms in on the top-right plot: we see that the transition lines are still slightly split.
Consistent with splitting due to $1/N_c$ corrections, the $N_c=3$ and $4$ lines are further separated than the $N_c=4$ and $5$ lines.

\subsection{Fermion Independence}

In QCD, observables typically have some dependence on the quark mass $m_q$.
However, we expect the theory to act increasingly like pure gauge theory as $N_c$ becomes large.
Thus, the value of any observable should lose its $m_q$ dependence as $N_c \rightarrow \infty$.
A slightly stronger prediction is that any observable should converge to its pure gauge value, regardless of the value of $m_q$, as $N_c$ becomes large.
Consistent with these predictions, the slope of each transition line in the bottom panel of Fig.~\ref{fig:dyn-diag-collapse} appears to be growing steeper with increasing $N_c$ and thus less sensitive to $\kappa$.

\begin{figure}[htp]
	\center
	\includegraphics[width=\textwidth]{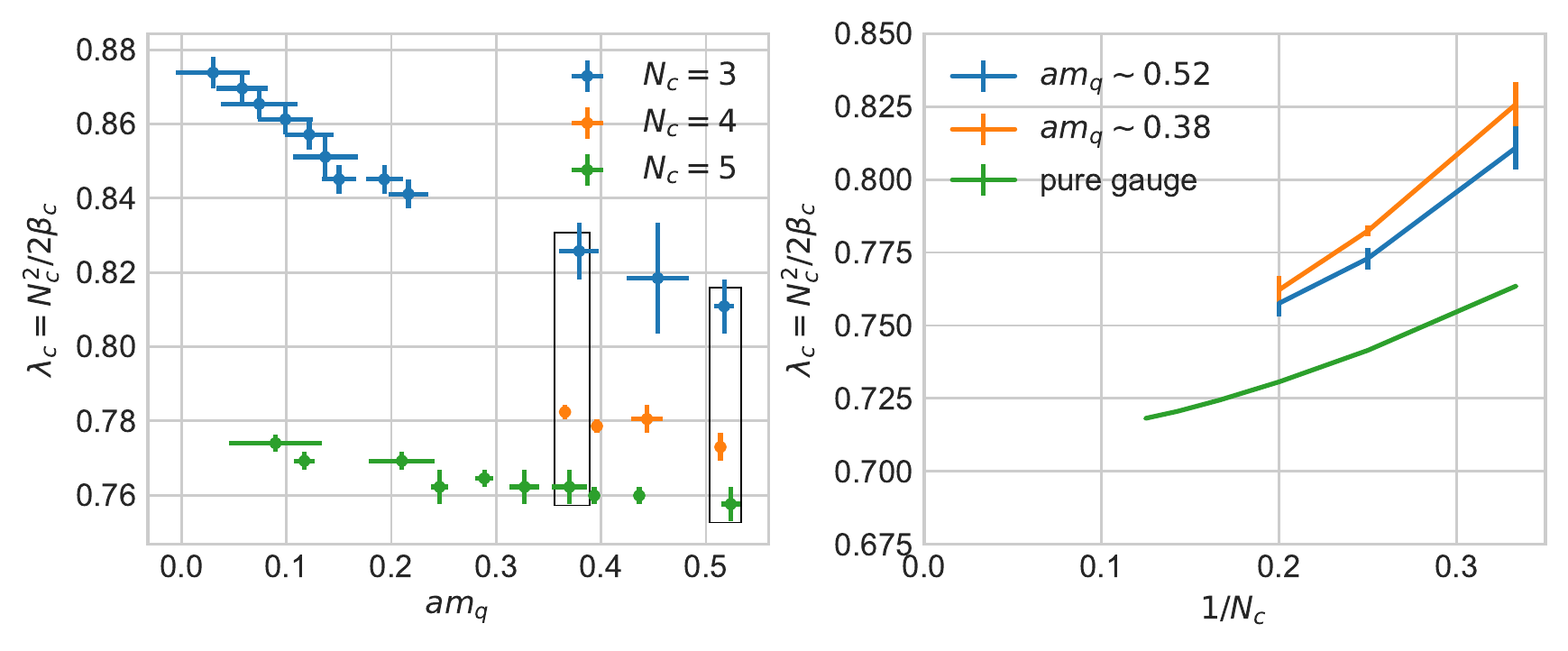}
	\caption{
		Each point corresponds to a fixed $\kappa$ in the indicated-$N_c$ phase diagram where the phase-ambiguous region is bracketed in $\beta$ by a confined and deconfined ensemble.
		The boxed ensembles at left are two slices that have roughly constant $am_q$ as $N_c$ is varied.
		These are plotted at right versus $1/N_c$ alongside pure-gauge data taken from Refs.~\protect\cite{Lucini:2012wq} and \protect\cite{Lucini:2003zr}.
		An apparent bulk transition blocks access to lighter $am_q$ for $N_c > 3$.
	}
	\label{fig:critical-thooft}
\end{figure}

We attempt to make this more quantitative in Fig.~\ref{fig:critical-thooft}.
On each phase diagram, we find fixed $\kappa$s where a confined ensemble and a deconfined ensemble bracket the phase-ambiguous region.
For each bracketing pair, we estimate $\beta_c$ and $am_q$ as the average between the two ensembles, with half the spread as the error.
From $\beta_c$ we compute the pseudocritical bare 't Hooft coupling $\lambda_c = N_c^2/2 \beta_c$, which has no leading $N_c$ dependence.
We plot $\lambda_c$ versus $a m_q$ in the left panel of Fig.~\ref{fig:critical-thooft}, where we see that the slope  indeed becomes shallower with increasing $N_c$.
The boxed ensembles in the left panel correspond to lines of roughly constant quark mass.
In the right panel, we plot $\lambda_c$ versus $1/N_c$ for these lines alongside the same curve for pure-gauge theory \cite{Lucini:2012wq,Lucini:2003zr}.
The two constant mass lines are similar, and could plausibly converge with the pure gauge line as $N_c \rightarrow \infty$.

\subsection{Pure-gauge-like Transition}

Pure gauge theory for $N_c \ge 3$ exhibits a first-order deconfinement transition \cite{Lucini:2012wq}, while in QCD, the transition at physical quark masses is a crossover \cite{Soltz:2015ula}.
If pure gauge theory is the $m_q \rightarrow \infty$ limit of QCD, then there must exist some critical quark mass $m_q^{PG}$ where the QCD-like crossover becomes a pure-gauge-like first-order transition.
We expect the theory to behave increasingly like pure gauge theory as we increase $N_c$, so we expect $m_q^{PG} \rightarrow 0$ as $N_c \rightarrow \infty$.
Meanwhile, stability analyses of the chiral transition \cite{Pisarski:1983ms, *Basile:2005hw} predict that it will be second order for $N_f=2$ in the chiral limit for any $N_c$.
It is ambiguous what will happen to the transition in the exact $m_q=0$ limit as $N_c \rightarrow \infty$.

Our current data set, with only one aspect ratio, is insufficient to perform a volume scaling analysis that would allow us to say something conclusive about the orders of the transitions that we observe.
However, we see no obvious indications of first-orderness anywhere in our data.
All ensembles we have generated have equilibrated quickly, without any tunneling events or other signs of metastability.
All observables we have examined are continuous across the transition.
The Polyakov loop at long flow time interpolates smoothly between nearly-zero in the confined phase and its maximum value in the deconfined phase, rather than exhibiting binary behavior.
All of this is different from what we observed in a previous study of a system with a strongly first-order transition \cite{Ayyar:2018ppa}.
If we are indeed observing crossovers, then we may bound $am_q^{PG} \gtrsim 0.5$.

\section{Outlook}
\label{sec:conclusion}

Enabled by automation, we can explore in many directions.
Phase diagrams for $N_t > 6$ will enable us to perform continuum $a \rightarrow 0$ extrapolations and move us away from the bulk transitions blocking exploration in $N_c > 3$.
Interpolating more finely near the transitions will permit us to look for peaks in susceptibilities, allowing us to determine the location of pseudocritical lines more quantitatively.
Different aspect ratios than $N_s/N_t=2$ will allow us to investigate finite volume effects in our data, which the Eguchi-Kawai volume reduction argument says should become smaller as $N_c$ increases \cite{Eguchi:1982nm}.
These would also allow for volume scaling analyses to more carefully determine transition order.
Zero-temperature data at bare parameters near the transitions will provide the lattice spacing and the AWI quark mass without contamination by finite-temperature effects.
If bulk transitions continue to be an issue at longer $N_t$, it may be worth considering improved actions.

Our investigation is currently limited to the line of constant $N_f=2$, but theories with other $N_f$ are also interesting.
For example, $N_f=3$ is another potentially-interesting limit of QCD.
Additionally, analytic predictions exist for the Veneziano limit $N_c \rightarrow \infty$ holding $N_f/N_c$ fixed {\cite{Veneziano:1974fa,*Veneziano:1976wm}}; exploring the grid of low-lying $N_c$ and $N_f$ would allow us to extrapolate to this limit for various $N_f/N_c$.

\begin{acknowledgments}
This work was supported in part by the U.S. Department of Energy under 
grant DE-SC0010005.
  Brookhaven National Laboratory is supported by the U.S. Department of 
Energy under contract DE-SC0012704.
\end{acknowledgments}

\bibliographystyle{JHEP}
\bibliography{lattice2018}

\end{document}